\documentclass[aps,prl,twocolumn,showpacs,floatfix]{revtex4}
\usepackage[dvips]{epsfig}
\begin{document}
{\small To be published in International Reviews in Physical
Chemistry, Vol.\ 25, issue 4 (Oct-Dec 2006) \hfill}
\newcommand{\App}{A^{\prime\prime}}
\title{Molecule formation in ultracold atomic gases}
\author{Jeremy M. Hutson}
\affiliation{Department of Chemistry, University of Durham, South
Road, Durham, DH1~3LE, England}
\author{Pavel Sold\'{a}n}
\affiliation{Doppler Institute,Department of Physics, Faculty of
Nuclear Sciences and Physical Engineering, Czech Technical
University, B\v{r}ehov\'{a} 7, 115 19 Praha 1, Czech Republic}

\date{21 July 2006}

\begin{abstract}
This review describes recent experimental and theoretical advances
in forming molecules in ultracold gases of trapped alkali metal
atoms, both by magnetic tuning through Feshbach resonances and by
photoassociation. Molecular Bose-Einstein condensation of
long-range states of both boson dimers and fermion dimers was
achieved in 2002-3. Condensates of boson dimers were found to be
short-lived, but long-lived condensates of fermion dimers have
been produced. Signatures of triatomic and tetraatomic molecules
have recently been observed. Both homonuclear and heteronuclear
molecules have been formed by photoassociation, mostly in very
high vibrational levels. Recent attempts to produce ultracold
molecules in short-range states (low vibrational levels) are
described. Experimental and theoretical work on collisions of
ultracold molecules is discussed.
\end{abstract}
\pacs{33.80.Ps,34.20.Mq,34.50.Lf,34.50.-s,03.75.Nt,03.75.Ss}

\maketitle

\tableofcontents

\section{Introduction}

The achievement of Bose-Einstein condensation in 1995 in dilute
gases of $^{87}$Rb \cite{Anderson:1995}, $^7$Li
\cite{Bradley:1995} and $^{23}$Na \cite{Davis:1995} revolutionised
atomic physics. Since that time, Bose-Einstein condensation has
been achieved for several other alkali metal species ($^{85}$Rb
\cite{Cornish:2000}, $^{41}$K \cite{Modugno:2001} and $^{133}$Cs
\cite{Weber:CsBEC:2003}) and a few other systems ($^1$H
\cite{Fried:1998}, metastable He \cite{Robert:2001,
DosSantos:2001}, $^{174}$Yb \cite{Takasu:2003} and $^{52}$Cr
\cite{Griesmaier:2005}). Intense effort has been devoted to the
study of the new properties of Bose-Einstein condensates
\cite{Cornell:2002, Ketterle:2002, Pethick:2002, Cornish:2003}.
%such as matter wave interference \cite{Andrews:1997}, atom lasers
%\cite{Mewes:1997, Bloch:1999}, vortices \cite{Matthews:1999},
%solitons \cite{Khaykovich:2002} and optical lattice effects
%\cite{Greiner:2002}.
The field was further broadened by the achievement of quantum
degeneracy in Fermi gases of $^{40}$K \cite{DeMarco:1999} and
$^6$Li \cite{Truscott:2001}, and ultracold fermionic quantum
matter has proved to exhibit a new range of novel properties.

Bose-Einstein condensation and Fermi degeneracy in dilute gases
typically require temperatures between 1 nK and 1 $\mu$K. However,
new quantum properties start to appear at temperatures around 1
mK, where de Broglie wavelengths become large compared to atomic
and molecular dimensions. Under these circumstances, collisions
are fully quantum-mechanical and are primarily sensitive to
long-range interactions. The region below 1 mK is generally
referred to as the {\it ultracold} regime.

Over the last few years, the focus of research in quantum matter
has shifted to the {\it control} of ultracold quantum systems. A
particularly important development has been the ability to form
and manipulate {\it molecules} in ultracold atomic gases.
Molecules have a much richer energy level structure than atoms,
and offer many new possibilities for quantum control. Perhaps most
importantly, {\it dipolar} molecules interact with one another
much more strongly and at longer range than atoms. Dipolar quantum
gases are predicted to exhibit more new features
\cite{Baranov:2002} and have possible applications in quantum
computing \cite{DeMille:2002}. Cold molecules also have
applications in high-precision measurement, and high-resolution
spectroscopy on cold molecules may allow the measurement of
fundamental physical properties such as the electric dipole moment
of the electron \cite{Hudson:2002}, the energy differences between
enantiomers (which result from parity violation) \cite{Quack:2005,
Crassous:2005} and the time-dependence of the fine-structure
constant \cite{Hudson:2006}.

There are two approaches to producing molecular quantum gases:
{\it direct} approaches, in which pre-existing molecules are
cooled from room temperature, and {\it indirect} approaches, in
which molecules are formed from precooled atoms. The direct
approaches have been reviewed previously \cite{Bethlem:2003} and
there are also several reviews that focus on the applications of
scattering theory to directly cooled molecules
\cite{Krems:IRPC:2005, Bodo:IRPC:2006, Weck:IRPC:2006}. The
purpose of this article is to review molecule formation in
ultracold quantum gases by {\it indirect} methods.

\section{Basic properties of ultracold atomic gases}

The physics of cooling and trapping atoms \cite{Adams:1997,
Chu:1998, Phillips:1998, Cohen-Tannoudji:1998} and of the
production and properties of Bose-Einstein condensates
\cite{Cornell:2002, Ketterle:2002, Pethick:2002, Cornish:2003} has
been reviewed many times. We will restrict ourselves here to a
brief discussion, focussing on aspects of the subject that are
essential to understanding the present review but may be
unfamiliar to readers with a background in physical chemistry
rather than atomic physics.

\subsection{Bosons and fermions}

All fundamental particles are either bosons or fermions. Bosons
have integer spin quantum numbers, while fermions have odd
half-integer spin quantum numbers. The fundamental difference is
encompassed by the {\em Pauli Principle}, which states that the
total wavefunction for a system must be symmetric with respect to
exchange of any pair of identical bosons, but antisymmetric with
respect to exchange of a pair of identical fermions. The most
important consequence of the Pauli Principle is the {\em Pauli
Exclusion Principle}, which states that no two fermions in the
same spin state can occupy the same spatial state.

Under circumstances where individual electrons cannot be
exchanged, atoms and molecules are {\em composite} bosons or
fermions. Any atom or molecule with an even number of nucleons and
electrons is a composite boson and any with an odd number is a
composite fermion. For the alkali metals, with an odd number of
electrons, isotopes with bosonic nuclei ($^6$Li, $^{40}$K) are
composite {\it fermions} and isotopes with fermionic nuclei
($^7$Li, $^{23}$Na, $^{41}$K, etc.) are composite {\it bosons}.

\subsection{Hyperfine structure}

\begin{figure} [htbp]
\begin{center}
\includegraphics[width=85mm]{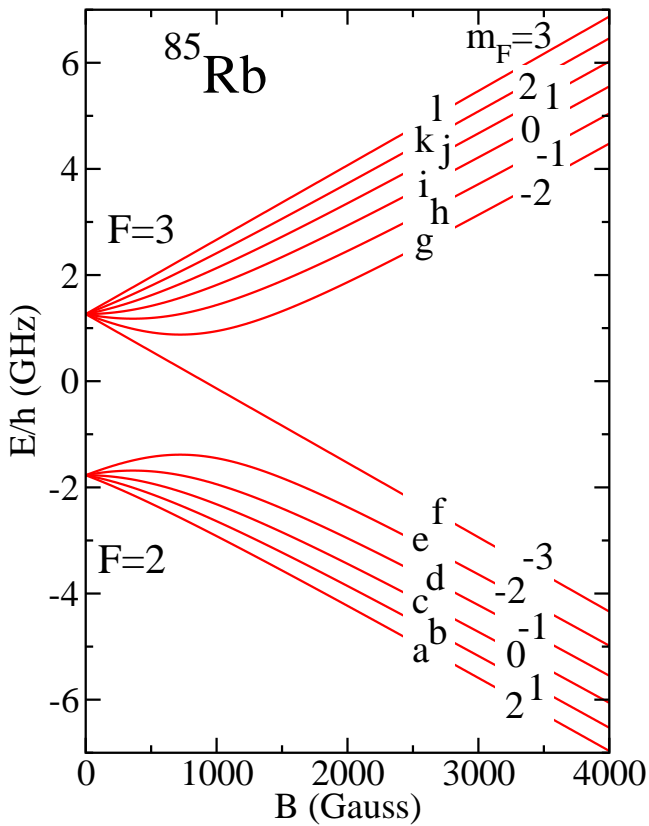}
\caption{Zeeman splitting of the hyperfine levels of $^{85}$Rb in
a magnetic field. $^{85}$Rb has $i=5/2$ and $f=2$ and 3. Figure
from Tiesinga \cite{Tiesinga:private:2006}.} \label{rb85-zeeman}
\end{center}
\end{figure}

An alkali metal atom with nuclear spin $i$ in its ground
electronic state ($^2$S$_{1/2}$) can have total angular momentum
$f=i\pm\frac{1}{2}$. In a magnetic field, the energy levels split
into components labelled by the projection quantum number $m_f$,
as shown for $^{85}$Rb in Fig.\ \ref{rb85-zeeman} (the Breit-Rabi
diagram). An atom for which the ground state has the lowest value
of $f$ is said to have {\it regular} hyperfine structure, and one
in which the order is reversed is said to have {\it inverted}
hyperfine structure. The only alkali metal atom important to the
present review that has inverted hyperfine structure is $^{40}$K.

The projection quantum number $m_f$ is a good quantum number at
any field, but $f$ is conserved only at zero field. At fields
above the avoided crossings in Fig.\ \ref{rb85-zeeman}, $f$ no
longer describes the character of the states. In this region the
nearly conserved quantities are the individual projections $m_s$
and $m_i$. This occurs at quite high field for $^{85}$Rb but at
much smaller fields for atoms with smaller hyperfine splittings
such as $^6$Li. Nevertheless, $(f,m_f)$ always provides a unique
{\it label} for a hyperfine state by following the curve back to
low field.

Molecules may have more than one nucleus with non-zero spin, and
have mechanical as well as electronic angular momentum. Their
energy level structure is thus considerably more complex than for
atoms. Nevertheless, it remains true that the zero-field levels
are characterised by a total angular momentum $f$, and that in a
field these are split into components with projection quantum
number $m_f$.

\subsection{Trapping and cooling}

Ultracold atoms and molecules would condense if they came into
contact with the walls of a vessel. It is therefore necessary to
trap them without physical walls. In most experiments, an atomic
beam is first decelerated using light pressure in a Zeeman slower
\cite{Phillips:1982}, which maintains the atoms in resonance as
they slow down. The slow atoms are then confined in a magnetic
\cite{Migdall:1985} or magneto-optical \cite{Raab:1987} trap.

Levels whose energy {\em increases} in a magnetic field (see Fig.\
\ref{rb85-zeeman}) are known as {\em low-field-seeking}, and those
whose energy {\em decreases} in a field are known as {\em
high-field-seeking}. A magnetic trap \cite{Migdall:1985} operates
by creating a local minimum in the magnetic field strength $B$, so
that atoms in low-field-seeking states are trapped. Since it is
not possible to create a local {\it maximum} in $B$ in free space,
high-field-seeking states cannot be trapped magnetically. For
alkali metal atoms, a magnetic trap has a typical depth of around
1 mK.

A magneto-optical trap (MOT) \cite{Raab:1987} uses a combination
of magnetic fields and Doppler cooling \cite{Chu:1985} to trap
atoms at considerably higher densities than is possible with
magnetic fields alone. A MOT can also cool atoms below the
theoretical limit of Doppler cooling by a mechanism known as
Sisyphus cooling \cite{Dalibard:1989}. For example, in early work
on $^{23}$Na, Lett {\em et al.}\ \cite{Lett:1988} observed a
temperature of 40 $\mu$K, compared to the 240 $\mu$K expected.
Even this is not by itself sufficient to achieve Bose-Einstein
condensation, and a final stage of cooling, often by evaporative
cooling \cite{Masuhara:1988}, is needed to achieve sub-$\mu$K
temperatures.

Ultracold atoms can also be confined in an optical dipole trap
\cite{Chu:1986}. Optical dipole traps have the advantage that all
magnetic sublevels can be trapped simultaneously. They rely on the
fact that the energy of an atom in an electric field oscillating
at frequency $\omega$ is $E=-\frac{1}{2} \alpha(\omega) F$, where
$F$ is the electric field strength. The frequency-dependent
polarizability $\alpha(\omega)$ is positive at low frequencies but
is enhanced (and changes sign) near absorption frequencies. An
optical dipole trap operates by creating an electric field {\em
maximum} in a region of strong laser radiation, and can be either
{\it near-resonant}, taking advantage of the enhancement in
$\alpha(\omega)$ near an absorption, or {\em far off-resonance}. A
far off-resonance trap (FORT) \cite{Miller:1993} causes less
heating but is much shallower than a near-resonant trap (sometimes
as little as 10 $\mu$K). An optical dipole trap in which the laser
frequency is so low that the polarizability is close to its static
value is referred to as a quasi-electrostatic trap (QUEST)
\cite{Takekoshi:1996}.

An optical dipole trap can confine molecules as well as atoms
\cite{Takekoshi:1998}. In addition, molecules with magnetic dipole
moments can be trapped magnetically \cite{Weinstein:CaH:1998,
Vanhaecke:2002}. Molecules with electric dipole moments can be
trapped by an analogous {\em electrostatic} approach
\cite{Bethlem:trap:2000}.

\subsection{Bose-Einstein condensation and Fermi degeneracy}

Most traps produce a trapping potential that is nearly harmonic
near the minimum. Because of this, the ``translational'' energy
spectrum of trapped atoms is not actually continuous, as is the
case for free particles. Instead, it is discretised by the
confinement in the trap: the energy level spacings $\hbar\omega$
are typically 10 to 1000 Hz, corresponding to 0.5 to 50 nK.

Bosons and fermions follow quite different quantum statistics:
Bose-Einstein and Fermi-Dirac statistics respectively. In a
Bose-Einstein condensate, nearly all the atoms are in the lowest
level in the trap (though there are always uncondensed atoms
coexisting with the condensate). In a harmonic trap, condensation
occurs at a critical temperature $T_{\rm c}$ given approximately
by $k_{\rm B}T_{\rm c}=0.94\hbar\omega N^{1/3}$, where $N$ is the
number of atoms \cite{Pethick:2002}. In a Fermi-degenerate gas, by
contrast, each level can accommodate only one atom in each spin
state ($f,m_f$), so that the system is characterised by full
occupation of trap levels up to the Fermi level, forming a ``Fermi
sea''. The Fermi temperature $T_{\rm F}$ is again proportional to
the level spacing, $k_{\rm B}T_{\rm F}=\hbar\omega (6N)^{1/3}$
\cite{Pethick:2002}. Both Bose-Einstein condensation and Fermi
degeneracy typically occur at temperatures below 1 $\mu$K.

\subsection{Scattering lengths}

The scattering wavefunction for collision of a pair of
structureless atoms is conveniently written
$\psi(r)=r^{-1}\chi(r)$. The radial wavefunction $\chi(r)$ obeys
the 1-dimensional Schr\"odinger equation,
\begin{equation}
\left[-\frac{\hbar^2}{2\mu}\frac{d^2}{dr^2}+V(r)-E\right]\chi(r)=0,
\end{equation}
where $E$ is the total energy, $V(r)$ is the effective potential
energy and $\mu$ is the reduced mass. For two atoms colliding with
zero kinetic energy, $V(r)\rightarrow E$ as $R\rightarrow\infty$,
so $\chi(r)$ has zero curvature in that region and becomes linear
in the interatomic distance $r$,
%At long range, the scattering wavefunction for a pair of atoms
%with zero kinetic energy becomes linear in the interatomic
%distance $r$,
\begin{equation} \chi(r) \sim r - a \quad \hbox{as} \quad
r\rightarrow\infty. \label{AsymptWaveFun}
\end{equation} The atom-atom interaction is characterised at the
simplest level by the {\it scattering length}, $a$, which is the
distance at which the continuation of the asymptotic straight line
(\ref{AsymptWaveFun}) crosses zero.

Many of the properties of a Bose-Einstein condensate depend only
on the scattering length and are unaffected by short-range
properties of the wavefunction such as the number of nodes. The
chemical potential $\mu_{\rm Bose}$ of a uniform Bose gas is
proportional to $a$, $\mu_{\rm Bose}=nU_0$, where $n$ is the
number density, $U_0=4\pi\hbar^2a/m$ and $m$ is the atomic mass
\cite{Pethick:2002}. A positive scattering length thus corresponds
to an interaction that is overall repulsive, while a negative
scattering length corresponds to an interaction that is overall
attractive. A large Bose-Einstein condensate can usually exist
only for positive scattering lengths; for negative scattering
lengths there is a limit to the size of condensate that can be
formed \cite{Kagan:1996, Bradley:1997}.

The scattering length is closely related to the energy of the
highest bound state of the atom-atom pair, $E_{\rm top}$. If
$E_{\rm top}$ is small and negative (a bound state just below
threshold), the scattering length is large and positive and is
related to the bound-state energy approximately by
\begin{equation} E_{\rm top} = \frac{-\hbar^2}{2\mu a^2}.
\label{eqetopa} \end{equation} The scattering length increases to
infinity as $E_{\rm top}$ approaches zero and reappears at large
negative values when $E_{\rm top}>0$.

The scattering length depends on the interaction potential and on
the reduced mass. It is thus different for pairs of atoms
interacting on singlet and triplet curves, and indeed for
different hyperfine states. It is also different for different
isotopic species.

\subsection{Feshbach resonances}

\begin{figure} [htbp]
\begin{center}
\includegraphics[width=85mm]{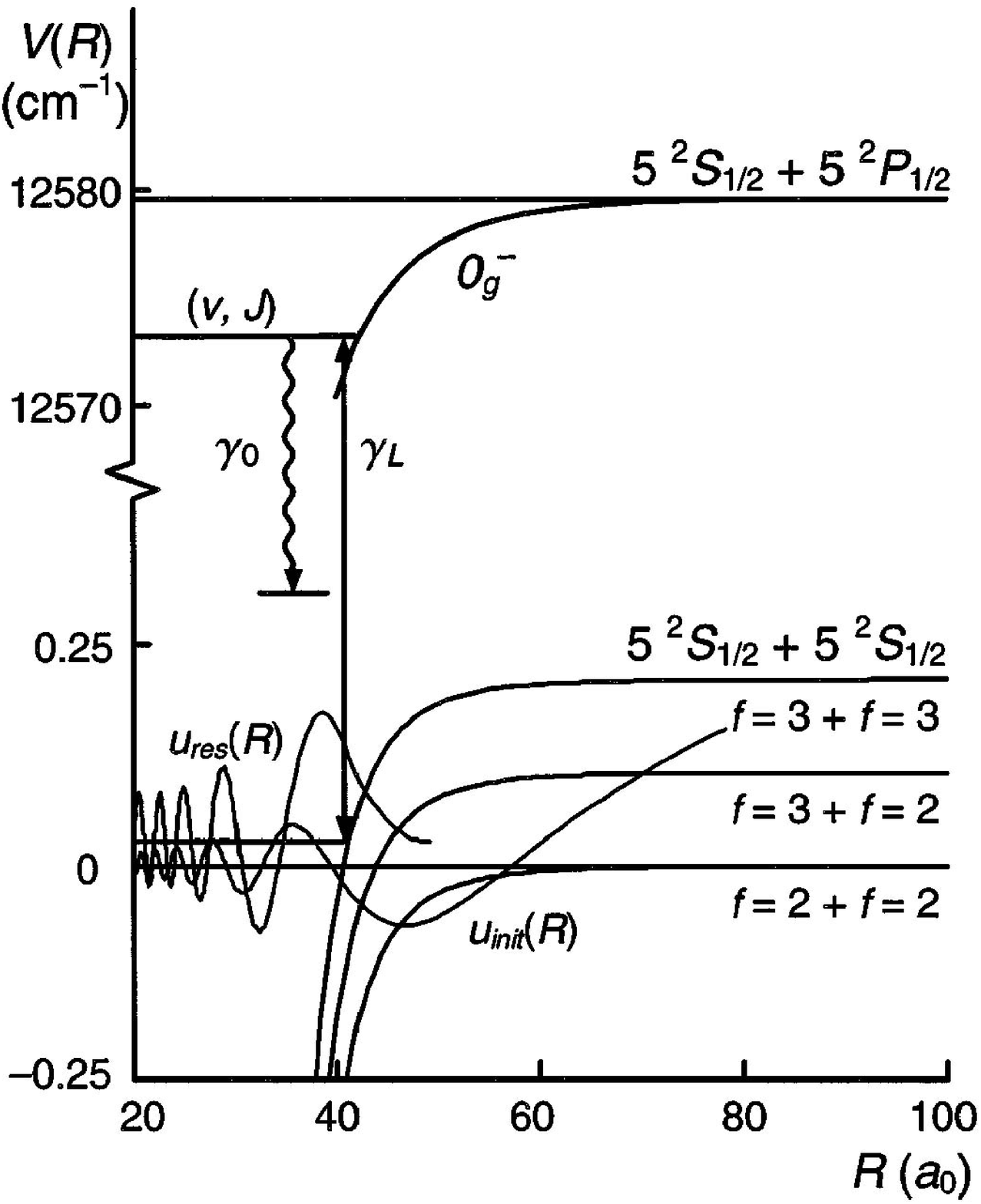}
\caption{Potential energy curves for $^{85}$Rb$_2$ (with nuclear
spin $i=5/2$) showing the 3 hyperfine thresholds. Also shown is a
zero-energy scattering wavefunction on the lowest curve
$(f_1,m_{f1};f_2,m_{f2})=(2,-2;2,-2)$ and a bound-state
wavefunction on the $(3,-2;3,-2)$ curve that is above the lowest
threshold and produces a Feshbach resonance. Reprinted with
permission from Courteille {\em et al.}\ \cite{Courteille:1998}.
Copyright 1998 by the American Physical Society.}
\label{feshbach-curves}
\end{center}
\end{figure}

A very important discovery was that the interactions between
ultracold atoms can be tuned using magnetic fields
\cite{Tiesinga:1993}. As described above, an atom with nuclear
spin $i$ in a $^2$S$_{1/2}$ state can have total angular momentum
$f=i\pm\frac{1}{2}$. When two such atoms interact, there are 3
closely-spaced thresholds corresponding to different combinations
of hyperfine states as shown in Fig.\ \ref{feshbach-curves}, and
there are sets of vibrational levels correlating with each
threshold. Some of the high-lying vibrational levels of the upper
curves can lie above the lower thresholds. Vibrational levels
embedded in a continuum are quasibound and produce Feshbach
resonances \cite{Feshbach:1958}. In zero field, such resonances
are characterised by their energy $E_{\rm res}$ and width $\Gamma$
(in energy space).

Each combination of atomic quantum numbers produces a {\em
channel}, and in general there are several channels correlating
with each threshold (corresponding to different values of
projection quantum numbers $m_f$ and values of the quantum numbers
$l$ and $m_l$ that describe mechanical rotation of the atoms about
one another). Each channel has its own potential energy curve,
though for simplicity only one curve is shown for each threshold
in Fig.\ \ref{feshbach-curves}. At energy $E$, each channel is
described as either {\em open} or {\em closed} (energetically
accessible or inaccessible at $R=\infty$). For example, at the
energy of the state shown in Fig.\ \ref{feshbach-curves}, channels
corresponding to the lowest threshold are open and the rest are
closed.

\begin{figure} [thbp]
\begin{center}
\includegraphics[width=85mm]{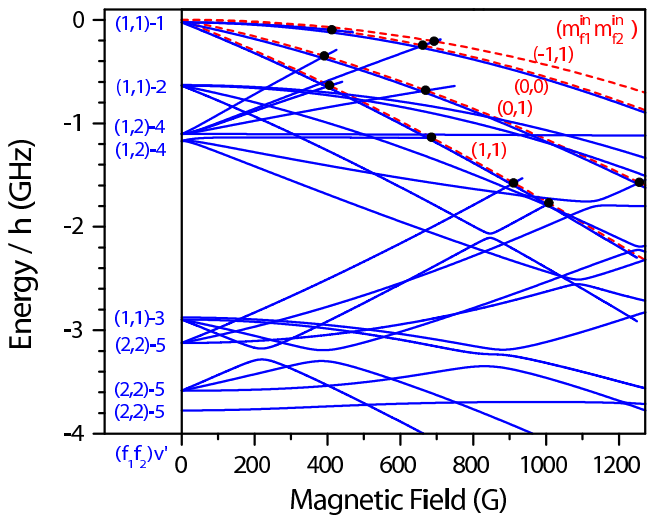}
\caption{Tuning of molecular levels (solid lines) and atomic
thresholds (dotted lines) for $^{87}$Rb$_2$ as a function of
magnetic field. Feshbach resonances occur at the points marked
with filled circles, where a molecular state crosses a threshold.
$^{87}$Rb has $i=3/2$, so there are higher thresholds
corresponding to $(f_1,f_2)=(1,2)$ and (2,2). Reprinted with
permission from Marte {\em et al.}\ \cite{Marte:2002}. Copyright
2002 by the American Physical Society.} \label{rb87-tuning}
\end{center}
\end{figure}

In a magnetic field, both the atomic levels (thresholds) and the
molecular levels split and shift as shown for $^{87}$Rb in Fig.\
\ref{rb87-tuning}. A high-lying vibrational level correlating with
one hyperfine state can often be tuned across a lower threshold
with an applied magnetic field. In work on ultracold gases, where
the collision energy is typically fixed at very near zero and a
magnetic field can be varied, the term ``Feshbach resonance" has
come to be applied to the behaviour of scattering properties as a
function of applied field as a state crosses threshold.

\begin{figure} [htbp]
\begin{center}
\includegraphics[width=85mm]{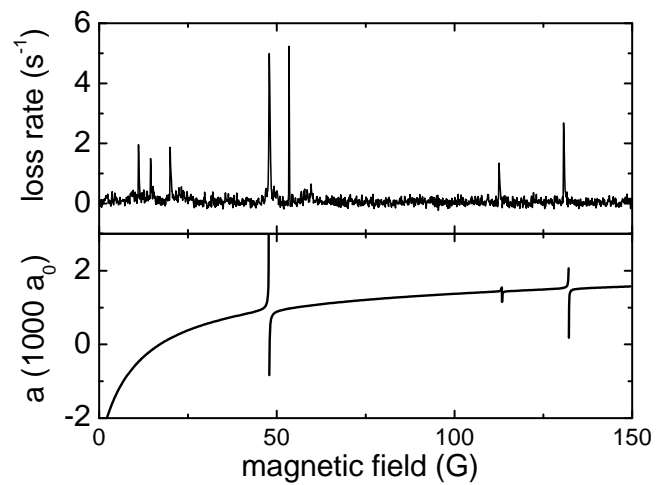}
\caption{Feshbach resonances in $^{133}$Cs collisions with
$(f,m_f)=(3,3)+(3,3)$ as a function of magnetic field . Lower
panel: scattering length based on theoretical parameters in ref.\
\onlinecite{Chin:cs2-fesh:2004}. Upper panel: loss rate for
radiative collisions \cite{Chin:2003}. Note that some of the
sharpest resonances are too narrow to see in the scattering length
but are still observed in the loss rates. Figure from Chin
\cite{Chin:private:2006}.} \label{cs-tuning-a+relax}
\end{center}
\end{figure}

Eq.\ \ref{eqetopa} applies even in the multichannel case, {\it
not} just when the scattering is governed by a single potential
curve. The scattering length thus has a pole whenever there is a
bound state at zero kinetic energy, as shown for $^{133}$Cs$_2$ in
the lower panel of Fig.\ \ref{cs-tuning-a+relax}. As a function of
magnetic field $B$, the scattering length in the vicinity of a
Feshbach resonance has the form \cite{Moerdijk:1995}
\begin{equation}
a(B)=a_{\rm bg}\left(1-\frac{\Delta B}{B-B_0}\right),
\label{a-res}
\end{equation}
where $a_{\rm bg}$ is the background scattering length, $B_0$ is
the resonance position (defined as the field at which $a$ is
infinite, which is not quite the same as the field at which the
bound state is at zero kinetic energy) and $\Delta B$ is related
to the width of the resonance $\Gamma$. In practice, it is of
course possible for resonances to overlap or for a sharp resonance
to occur in the wings of a broad one. For example, the strong
field-dependence of the scattering length for $^{133}$Cs$_2$ at
low magnetic field shown in Fig.\ \ref{cs-tuning-a+relax} can be
interpreted in terms of a broad Feshbach resonance at $B_0=-8$ G
\cite{Weber:2003}.

The elastic cross section $\sigma(k)$ for identical bosons at
kinetic energy $E_{\rm kin}=\hbar^2 k^2/2\mu$ is approximately
given by \cite{Joachain:1975}
\begin{equation} \sigma(k) = \frac{8\pi a^2}{1+(ka)^2} +
\ldots, \label{eqsigmaa} \end{equation} so that $\sigma(k)$ passes
through a peak of height $8\pi/k^2$ (corresponding to $a=\infty$)
at a resonance. Other collisional properties also show sharp
features. For example, for $^{133}$Cs the radiative loss rate
(which is due to atom pairs temporarily excited to the excited
electronic state) exhibits sharp peaks as shown in the upper panel
of Fig.\ \ref{cs-tuning-a+relax}
\cite{Chin:2003,Chin:cs2-fesh:2004}.

Magnetic tuning of Feshbach resonances was first observed by
Inouye {\em et al.}\ \cite{Inouye:1998} and Stenger {\em et al.}\
\cite{Stenger:1999}, who detected enhanced loss rates from a
Bose-Einstein condensate of $^{23}$Na as a function of magnetic
field, and by Courteille {\em et al.}\ \cite{Courteille:1998}, who
observed enhanced photoassociation rates in the vicinity of a
Feshbach resonance in $^{85}$Rb. The loss rates were attributed
either to excitation of atoms to higher trap states during the
field ramp \cite{vanAbeelen:1999, Mies:2000} or to inelastic
collisions involving the transiently formed molecular state
\cite{Yurovsky:1999}. In experiments on $^{85}$Rb, Roberts {\em et
al.}\ \cite{Roberts:collapse:2001} and Donley {\em et al.}\
\cite{Donley:2001} demonstrated that tuning the scattering length
suddenly from positive to negative could cause controlled collapse
of a Bose-Einstein condensate. The collapse produces an explosion
of atoms from the condensate and has come to be known as a
Bosenova by analogy with astrophysical supernovae.

\section{Molecules formed by Feshbach resonance tuning}

\begin{figure} [htbp]
\begin{center}
\includegraphics[width=85mm]{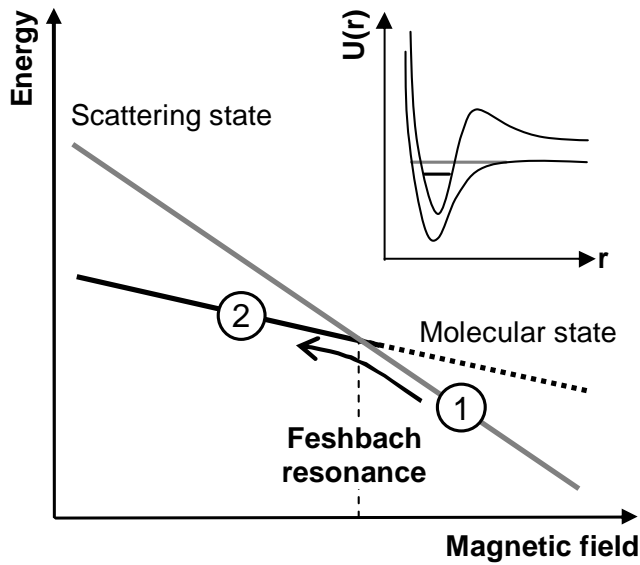}
\caption{The crossing of atomic (scattering) and molecular states
as a function of magnetic field, showing the avoided crossing and
the use of a field ramp to convert pairs of atoms into molecules.
Note that in this case the atomic and molecular states are both
high-field-seeking, so cannot be trapped magnetically. Reprinted
with permission from Herbig {\em et al.}\ \cite{Herbig:2003}.
Copyright 2003 by Macmillan Publishers Ltd.} \label{avoided}
\end{center}
\end{figure}

The possibility of using Feshbach resonances to create ultracold
molecules was first predicted in 1999 \cite{Timmermans:1999,
vanAbeelen:1999, Mies:2000}. As described above, a Feshbach
resonance occurs when a molecular level crosses an atomic
threshold as a function of the magnetic field $B$. In reality,
since the atomic and molecular states are coupled, there is an
avoided crossing. Thus if the magnetic field is tuned across the
resonance, slowly enough to follow the avoided crossing
adiabatically, pairs of trapped {\it atoms} can be converted into
trapped {\it molecules} as shown in Fig.\ \ref{avoided}. Since no
kinetic energy is created in the process, the molecules are
created with essentially the same energy as the atoms.

The actual arrangement of states in a Feshbach resonance is
sometimes the mirror image of that shown in Fig.\ \ref{avoided}.
In the following discussion, the side of the resonance where
molecules are lower in energy than atoms will be referred to as
the {\it molecular side} of the resonance.
%The state that is molecular in character (the grey line in Fig.\
%\ref{avoided} will be referred to as the {\it molecular curve}.

\subsection{Dimers of bosonic atoms}

The first signatures of trapped molecules produced by Feshbach
resonance tuning were observed by Donley {\em et al.}\
\cite{Donley:2002} in 2002. They worked with magnetically trapped
$^{85}$Rb, so were restricted to low-field-seeking states.
$^{85}$Rb has nuclear spin $i=5/2$, and Donley {\em et al.}\
worked with trapped atoms in the $(f,m_f)=(2,-2)$ state, which is
not the lowest state in a magnetic field (see Fig.\
\ref{rb85-zeeman}). In this case the energy levels are the mirror
image of those shown in Fig. \ref{avoided}, with the molecular
state below the atomic state on the {\it high-field} side of the
resonance.

The experiment of Donley {\it et al.}\ was a little more
complicated than a simple ramp over the resonance as shown in
Fig.\ \ref{avoided}. Since the background scattering length for
this system is negative, $a(B)$ is negative on the atomic side
even far from the resonance. For this reason it was not possible
to approach the resonance from the atomic side, because the
condensate collapsed \cite{Roberts:collapse:2001, Donley:2001}.
Donley {\em et al.}\ therefore used a magnetic field profile that
approached the Feshbach resonance from the {\em molecular} side
but did not actually cross it. The fast field ramp mixed the
atomic and molecular states, and Donley {\em et al.}\ observed
quantum beats (Ramsey fringes) between trapped atoms and
molecules.

\begin{figure} [htbp]
\begin{center}
\includegraphics[width=85mm]{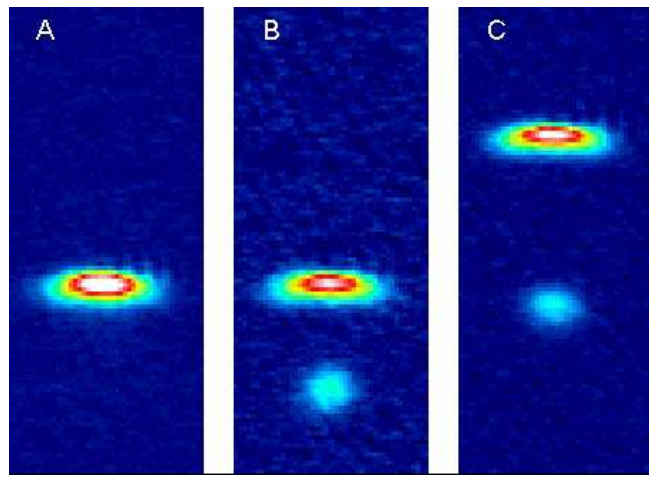}
\caption{Formation of a quantum gas of $^{133}$Cs$_2$ molecules.
A: magnetically levitated atomic BEC; B: levitated atomic BEC with
falling molecular cloud below; C: levitated molecular cloud with
rising atomic BEC above. Reprinted with permission from Herbig
{\em et al.}\ \cite{Herbig:2003}. Copyright 2003 by Macmillan
Publishers Ltd.} \label{Cs2BEC}
\end{center}
\end{figure}

Over the following two years, the technique was developed by
several groups and extended to other bosonic systems:
$^{133}$Cs$_2$ \cite{Herbig:2003}, $^{87}$Rb$_2$
\cite{Durr:mol87Rb:2004} and $^{23}$Na$_2$ \cite{Xu:2003}. In all
these experiments atoms were prepared in the {\it lowest}
hyperfine state in a magnetic field, with $f=i-1/2$ and $m_f=f$.
Such states cannot be trapped magnetically (because they are
high-field-seeking), so the atoms were confined in optical traps.
Molecules were created by sweeping the magnetic field {\em across}
a resonance as shown in Fig.\ \ref{avoided}. The molecules were
separated from the remaining atoms either magnetically
\cite{Herbig:2003, Durr:mol87Rb:2004} or by using a laser resonant
with the atoms but not the molecules to push the atoms out of the
trap \cite{Xu:2003}. The molecules were then detected and imaged
by converting them back to atoms with a reverse field sweep. Fig.\
\ref{Cs2BEC} shows images of the Cs$_2$ molecular cloud created by
Herbig {\em et al.}\ \cite{Herbig:2003}.

In all these experiments involving bosonic atoms it was found that
the molecules were lost from the trap within a few milliseconds.
The fast trap loss was attributed to atom-diatom collisions. The
molecules are formed in a very highly excited state, often the
highest vibrational state that exists in the potential well. There
are always many lower-lying states, and even near dissociation the
vibrational spacing is large compared to the depth of the trap.
Thus inelastic atom-molecule collisions,
\begin{equation} \hbox{M}_2(v) + \hbox{M} \longrightarrow
\hbox{M}_2(v^\prime<v) + \hbox{M}, \end{equation} or
molecule-molecule collisions,
\begin{equation} \hbox{M}_2(v) + \hbox{M}_2(v) \longrightarrow
\hbox{M}_2(v^\prime<v) + \hbox{M}_2 (v^{\prime\prime}\le v),
\end{equation} always release enough kinetic energy to eject both
collision partners from the trap. It should be noted that the
molecules are not {\it destroyed} in such collisions, but they are
lost from the trap and are no longer ultracold.

As remarked above, the molecules produced by Donley {\em et al.}\
\cite{Donley:2002} are formed from $^{85}$Rb atoms in the
$(f,m_f)=(2,-2)$ state, which is not the lowest state in a
magnetic field. They are thus not strictly bound, and can
dissociate spontaneously without collisions to form atoms in
lower-lying hyperfine states. This has been studied experimentally
by Thompson {\em et al.}\ \cite{Thompson:spont:2005} and
theoretically by K\"ohler {\em et al.}\ \cite{Kohler:2005}.
Thompson {\em et al.}\ \cite{Thompson:spont:2005} adapted the
experiment of Donley {\em et al.}\ \cite{Donley:2002} to create
$^{85}$Rb$_2$ molecules using a magnetic field sweep through the
Feshbach resonance without holding the atomic condensate at $a<0$
for long enough for it to collapse. As in the earlier experiments,
the molecules were formed in states above the lowest hyperfine
threshold. Even at densities where collisional loss was very slow,
Thompson {\em et al.}\ found that the molecules decayed within 1
ms at magnetic fields far from resonance. However, close to
resonance, they were able to achieve lifetimes of tens of
milliseconds. It should be noted that this decay mechanism is not
applicable to the experiments on $^{133}$Cs$_2$, $^{87}$Rb$_2$ and
$^{23}$Na$_2$ \cite{Herbig:2003, Durr:mol87Rb:2004, Xu:2003},
where the molecules are formed in states that lie below the lowest
atomic threshold.

\begin{figure} [htbp]
\begin{center}
\includegraphics[width=85mm]{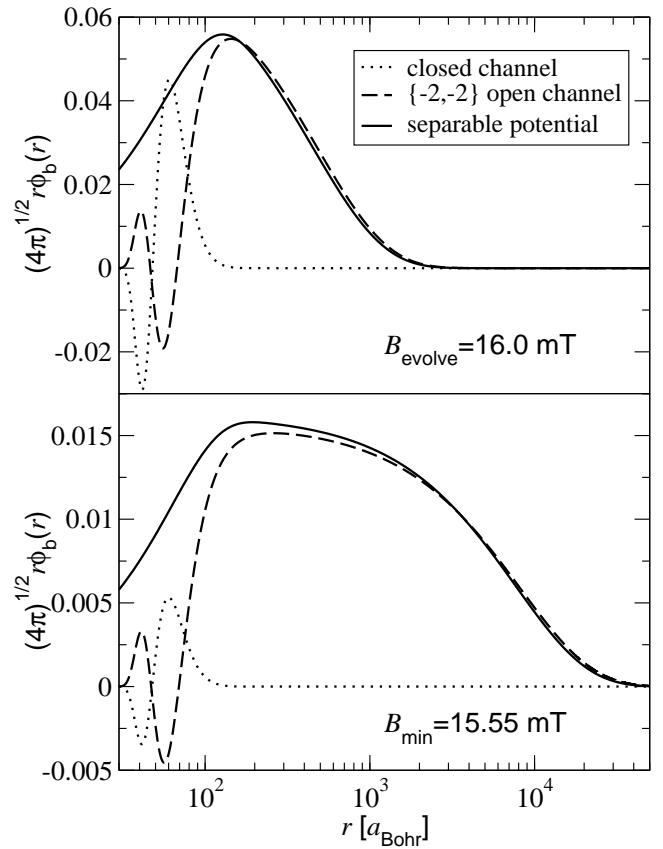}
\caption{Bound-state wavefunctions for a model of $^{85}$Rb$_2$
close to a Feshbach resonance ($B=15.55$ mT, $a=7800\,a_0$) and
slightly further away ($B=16.0$ mT, $a=521\,a_0$). Reprinted with
permission from K\"ohler {\em et al.}\ \cite{Kohler:2003}.
Copyright 2003 by the American Physical Society.}
\label{longrange}
\end{center}
\end{figure}

Molecules formed by Feshbach resonance tuning are very large. Even
the closed-channel part of the wavefunction corresponds to a
molecule in a very high vibrational state, as shown in Fig.\
\ref{feshbach-curves}; the wavefunction peaks near the outer
turning point, which can be at distances approaching $R=100\ a_0$.
However, even this underestimates the size of Feshbach molecules.
K\"ohler {\em et al.}\ \cite{Kohler:2003} have investigated a
model of the Feshbach resonance used to produce $^{85}$Rb$_2$ by
Donley {\em et al.}\ \cite{Donley:2002}, and have shown that for
magnetic fields near the resonance the molecular wavefunction is
dominated by the {\em open} (atomic) channel and has a
wavefunction that dies off as $\exp(-r/a)$ at long range; the mean
internuclear distance is on the order of $a/2$, which near a
resonance can be several thousand $a_0$. An example of this is
shown in Fig.\ \ref{longrange}.

\subsection{Dimers of fermionic atoms}

In parallel to the work on boson dimers, molecules were created
from pairs of {\it fermionic} atoms by tuning through Feshbach
resonances. Collision rates for two identical fermions are
suppressed because s-wave scattering (partial wave $l=0$) is
forbidden. Accordingly, the fermion work focussed on pairs of
atoms in different spin states, again in optical traps. Molecule
formation was first achieved for $^{40}$K$_2$ by Regal {\em et
al.}\ \cite{Regal:40K2:2003}. $^{40}$K has nuclear spin $i=4$ and
an inverted hyperfine structure, so that the atomic ground state
in a magnetic field has $(f,m_f)=(9/2,-9/2)$. Regal {\em et al.}\
used a mixture of $(9/2,-9/2)$ and $(9/2,-5/2)$ atoms and formed
molecules by magnetic tuning through a Feshbach resonance. Once
again the molecules were found to be short-lived ($\tau\approx 1$
ms). The lifetime was attributed to vibrationally inelastic
collisions \cite{Regal:40K2:2003}, though spontaneous dissociation
by spin relaxation is also possible in this case.

A major experimental breakthrough came in mid-2003, when four
groups \cite{Strecker:2003, Cubizolles:2003, Jochim:Li2pure:2003,
Regal:lifetime:2004} independently reported within a period of 6
weeks that fermion dimers can be remarkably stable to collisions
when the atom-atom scattering length is tuned to a large positive
value. $^6$Li has nuclear spin $i=1$, and Strecker {\em et al.}\
\cite{Strecker:2003} and Cubizolles {\em et al.}\
\cite{Cubizolles:2003} prepared $^6$Li$_2$ molecules by Feshbach
resonance tuning in mixtures of the lowest two spin states,
correlating with $(f,m_f)=(1/2,1/2)$ and $(1/2,-1/2)$ at low
field. Both groups showed that the molecules remained trapped for
1 s or more before being dissociated by a reverse magnetic field
sweep. Jochim {\em et al.}\ \cite{Jochim:Li2pure:2003} prepared
$^6$Li$_2$ molecules by a different method, taking advantage of
the dramatically enhanced 3-body recombination rate near a
Feshbach resonance, and observed comparable lifetimes. Cubizolles
{\em et al.}\ \cite{Cubizolles:2003} and Jochim {\em et al.}\
\cite{Jochim:Li2pure:2003} showed that the lifetime was
particularly large close to the resonance, where the scattering
length is large and positive. Regal {\em et al.}\
\cite{Regal:lifetime:2004} carried out analogous experiments on
$^{40}$K$_2$, both for the spin states involved in their earlier
experiments \cite{Regal:40K2:2003} and for molecules formed from
$(f,m_f)=(9/2,-9/2)$ and $(9/2,-7/2)$. They confirmed the fast
decay far from resonance, but showed that for large positive
scattering lengths the lifetime was dramatically enhanced.

\begin{figure} [htbp]
\begin{center}
\includegraphics[width=85mm]{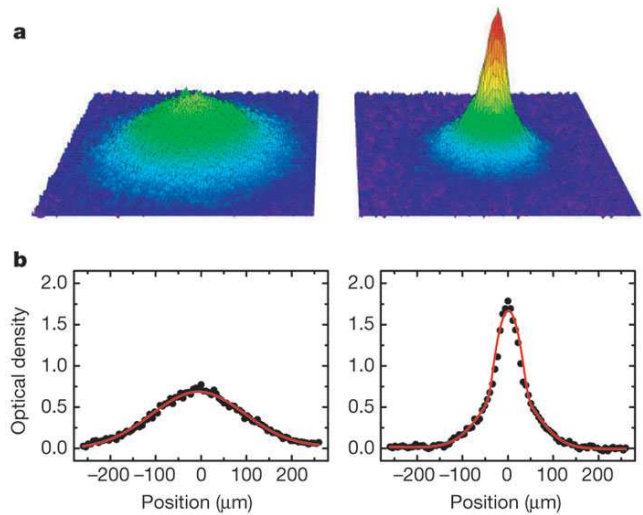}
\caption{Images of a molecular cloud of $^{40}$K$_2$ after 20 ms
of free expansion, above and below the critical temperature for
Bose-Einstein condensation. The condensed cloud (right) shows the
tight spatial peak characteristic of a condensate. Reprinted with
permission from Greiner {\em et al.}\ \cite{Greiner:2003}.
Copyright 2003 by [check organisation]} \label{k2bec}
\end{center}
\end{figure}

By the end of 2003, three different groups had succeeded in
creating long-lived molecular Bose-Einstein condensates of fermion
dimers. Jochim {\em et al.}\ \cite{Jochim:Li2BEC:2003} and
Zwierlein {\em et al.}\ \cite{Zwierlein:2003} achieved this by
evaporative cooling combined with 3-body recombination in a mixed
gas of $^6$Li in its $(f,m_f)=(1/2,1/2)$ and $(1/2,-1/2)$ states,
held at large positive scattering length close to a Feshbach
resonance. Under these circumstances 3-body recombination to form
molecules is thermodynamically favourable and the molecules are
long-lived. Greiner {\em et al.}\ \cite{Greiner:2003} formed
$^{40}$K$_2$ from a very cold Fermi gas of $^{40}$K$_2$ atoms in
their $(f,m_f)=(9/2,-9/2)$ and $(9/2,-7/2)$ states, using a
Feshbach ramp so slow that thermal equilibrium was maintained
during the field sweep. All three groups observed the sudden
appearance of a sharp spatial peak in the density with decreasing
temperature, as shown in Fig.\ \ref{k2bec} for $^{40}$K$_2$. This
is widely regarded as the ``smoking gun" of Bose-Einstein
condensation \cite{Anderson:1995, Davis:1995}.

Petrov {\em et al.}\ \cite{Petrov:2004, Petrov:suppress:2005}
analysed the stability of fermion dimers in terms of the
long-range form of the wavefunction. In the case where the
atom-atom scattering length $a$ is much larger than the range of
the atom-atom potential $r_e$, they showed that both the
atom-molecule and molecule-molecule inelastic collision rates are
suppressed by Fermi statistics. However, their derivation applies
only to molecules that are in long-range states, with a
wavefunction that depends on the scattering length, $\chi(r) \sim
\exp(-r/a)$. As will be discussed in more detail below, Cvita\v s
{\em et al.}\ \cite{Cvitas:bosefermi:2005} have shown
computationally that there is no systematic suppression of the
atom-molecule inelastic rate for fermion dimers in low-lying
vibrational levels, even when $a$ is large and positive.

A major reason for interest in fermion dimers is that they provide
tunable models for studying problems in condensed matter physics
such as the origin of superfluidity and superconductivity.
According to the widely accepted theory of Bardeen, Cooper and
Schrieffer (BCS) \cite{Bardeen:1957}, superconductivity occurs
because pairs of electrons are composite bosons and can therefore
condense. However, the pairs of electrons involved in
superconductivity (Cooper pairs) are much larger than the mean
separation between electrons. This is different from the usual
regime of molecular Bose-Einstein condensation (BEC), because far
from a resonance a fermion dimer is relatively strongly bound and
is small compared to the typical separation between unbound atoms.
However, close to resonance the ``size" of the molecules increases
and can become comparable to the atom-atom spacing. Under these
circumstances the ``molecules" are interpenetrating and lose their
identity in very much the same way as Cooper pairs of electrons.
The beauty of the fermionic atom systems is that this transition
can be followed as a function of magnetic field.

The BEC-BCS crossover between a molecular Bose-Einstein condensate
and a condensate of atomic Cooper pairs has been studied
extensively. The BCS regime is an intrinsically many-body regime
in which 2-body theories cannot be expected to work. At the centre
of a resonance the 2-body scattering length is infinite but the
pair size in a many-body system is limited to the atom-atom
spacing.
%Remarkably, this asymptotic behavior is
%predicted to be universal (independent of the details of the
%atom-atom potential). This universality was experimentally
%verified in 2002 \cite{O'Hara:2002, Bourdel:2003}.
Beyond the resonance, where the scattering length is negative,
%the fluctuations between the bound and continuum states continue
%to bind fermions into Cooper pairs, but
the binding forces are much smaller and the pair size becomes much
larger then the interparticle spacing. Bartenstein {\em et al.}\
\cite{Bartenstein:crossover:2004} showed that the transition
between the two regimes can be achieved smoothly and reversibly.
Regal {\em et al.}\ \cite{Regal:res-cond:2004, Regal:2005} and
Zwierlein {\em et al.}\ \cite{Zwierlein:2004} demonstrated
Bose-Einstein condensation of Cooper pairs in the BCS regime and
measured the atomic momentum distribution in the BCS region. Chin
{\em et al.}\ \cite{Chin:pairgap:2004} and Greiner {\em et al.}\
\cite{Greiner:2005} observed a pairing gap characteristic of
superfluidity. Partridge {\em et al.}\ \cite{Partridge:2005} have
measured the closed-channel component of the pair wavefunction on
both sides of a resonance and shown that, though small, it
persists on the BCS side of the resonance, contrary to the
predictions of 2-body theory.

Other studies have focussed on signatures of superfluidity rather
than Cooper pairing. Kinast {\em et al.}\ \cite{Kinast:2004} and
Bartenstein {\em et al.}\ \cite{Bartenstein:collective:2004}
observed collective oscillations in a strongly interacting Fermi
gas that suggest superfluid behaviour in the BCS regime, while
Kinast {\em et al.}\ \cite{Kinast:2005} measured the heat capacity
of such a gas. Most recently, Zwierlein {\em et al.}\
\cite{Zwierlein:2005} provided conclusive evidence of
superfluidity by observing arrays of vortices on both the BEC and
BCS sides of a resonance in $^6$Li$_2$.

\subsection{Heteronuclear Feshbach resonances}

It is possible to trap two different alkali metal species
simultaneously, and magnetic Feshbach resonances have been
observed in RbK \cite{Inouye:2004, Ferlaino:2006} and LiNa
\cite{Stan:2004}. There is little doubt that tuning through such
resonances will soon be used to form heteronuclear molecules.
However, it should be noted that neutral heteronuclear molecules
in long-range states do {\it not} have significant dipole moments:
typical values are less than 0.3 D at $R=15\ a_0$
\cite{Aymar:2005} and decay as $D_7 R^{-7}$ at long range
\cite{Brown:1973}.

\subsection{Triatomic and larger molecules}

It is in principle possible to form molecules larger than
diatomic, either by direct association from atoms or by
association of smaller molecules. Chin {\em et al.}\
\cite{Chin:2005} have formed Cs$_2$ molecules by Feshbach
resonance tuning and then separated out the remaining atoms
magnetically. They observed field-dependent resonances in the
inelastic loss rates that they attributed to Cs$_4$ bound states
near the molecular scattering threshold.

\begin{figure} [htbp]
\begin{center}
\includegraphics[width=85mm]{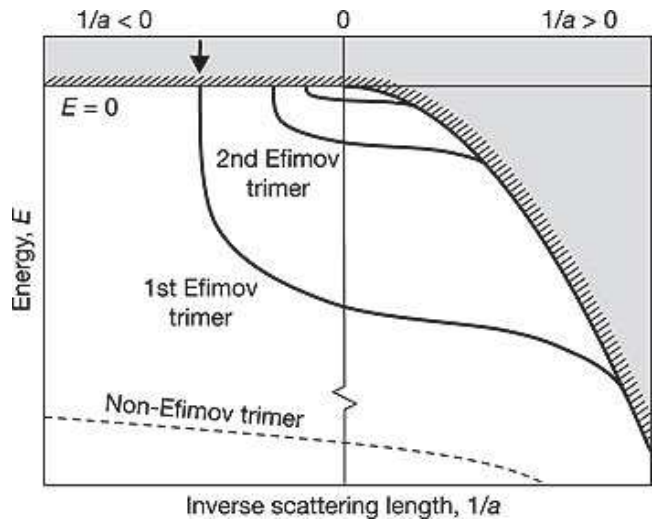}
\caption{Appearance of Efimov trimer states, showing how they
intersect the threshold for 3 separated atoms as a function of
scattering length. The shaded area shows the scattering continuum
for 3 atoms ($a<0$) and for atom + diatom ($a>0$). The
characteristic factor of 22.7 in $a$ has been reduced to 2 for the
purpose of illustration. Reproduced from Kraemer {\em et al.}\
\cite{Kraemer:2006}.} \label{efimov}
\end{center}
\end{figure}

For three atoms there is the intriguing prospect of forming Efimov
states \cite{Efimov:1970}, which are long-range trimer states that
exist even when the corresponding dimers are unbound. Indeed,
Efimov showed that, if the pair potential has exactly one bound
state at zero energy (corresponding to an infinite scattering
length), there are an infinite number of such trimer states. The
helium trimer is predicted to have one Efimov state
\cite{Lim:1977, Esry:1996}, but it has not yet proved possible to
observe this experimentally. However, alkali metal atoms with
tunable interactions offer new possibilities for a slightly
different type of Efimov state, which occurs whenever the
scattering length is large but is complicated by the existence of
a large number of deeply bound states. Braaten and Hammer
\cite{Braaten:2001} and Nielsen {\em et al.}\ \cite{Nielsen:2002}
have shown that Efimov states will cause resonant enhancements in
3-body recombination rates at characteristic values of the
scattering length that differ by successive factors of 22.7. The
characteristic dependence of energies on scattering length is
shown in Fig.\ \ref{efimov}. Very recently, Kraemer {\em et al.}\
\cite{Kraemer:2006} have measured trap loss in an ultracold gas of
Cs atoms as a function of scattering length, and observed a peak
that they attribute to resonance between an Efimov state of the
trimer and the threshold for 3 separated atoms.

\section{Molecules formed by photoassociation}

\begin{figure} [htbp]
\begin{center}
\includegraphics[width=85mm]{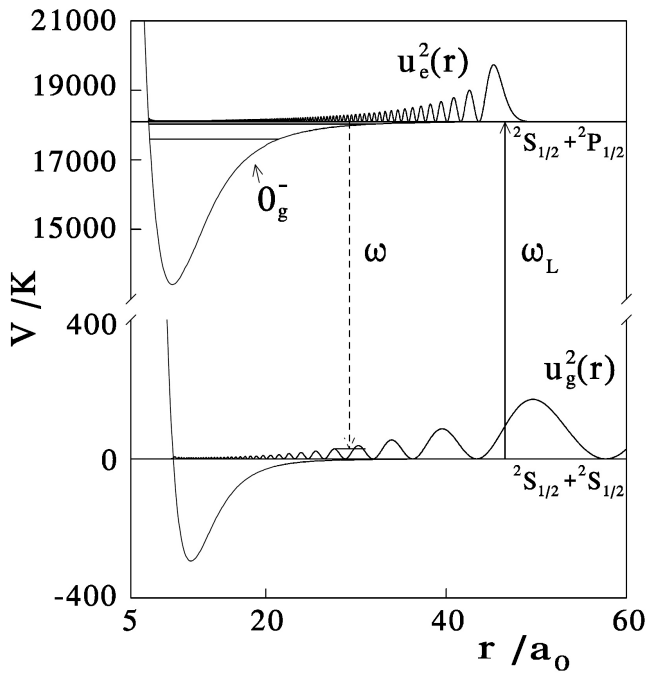}
\caption{Potential energy curves and (squares of) vibrational
wavefunctions for photoassociation of Rb$_2$. Reprinted with
permission from Boesten {\em et al.}\ \cite{Boesten:1999}.
Copyright 1999 by Institute of Physics.} \label{photoass-2colour}
\end{center}
\end{figure}

Dimers can also be formed in cold atomic gases by
photoassociation, as predicted by Thorsheim {\em et al.}\ in 1987
\cite{Thorsheim:1987}. Developments up to 1999 were reviewed by
Stwalley and Wang \cite{Stwalley:1999} and more recent work by
Jones {\em et al.}\ \cite{Jones:RMP:2006}. The early work focussed
on 1-photon photoassociation spectroscopy, forming molecules in
electronically excited states. More recently, however, it has
become possible to form ultracold molecules in their electronic
ground states by 2-photon processes as shown in Fig.\
\ref{photoass-2colour}. This was first achieved by Fioretti {\em
et al.}\ \cite{Fioretti:1998}, who photoassociated Cs$_2$ to an
excited $0_g^-$ electronic state with a double minimum
\cite{Fioretti:1999} and observed ultracold molecules formed by
spontaneous emission to the lowest triplet state. Nikolov {\em et
al.}\ \cite{Nikolov:1999} carried out a similar experiment to form
the $^1\Sigma_g^+$ ground state of K$_2$ and used resonant
two-colour ionization to show that the vibrational distribution
peaked at $v=36$, just over half way up the ground state potential
well. They subsequently \cite{Nikolov:2000} developed a 2-photon
excitation scheme via an excited $^1\Pi_u$ state that produced
molecules even further down the ground-state well. Optical
trapping of ground-state Cs$_2$ molecules was achieved by
Takekoshi {\em et al.}\ \cite{Takekoshi:1998} and magnetic
trapping by Vanhaecke {\em et al.}\ \cite{Vanhaecke:2002}.

\subsection{Photoassociation in Bose-Einstein Condensates}

An additional degree of control can be introduced by using
stimulated Raman adiabatic passage (STIRAP) \cite{Gaubatz:1990},
in which a second laser detuned from the excitation laser brings
the molecules down to bound levels of the ground electronic state
\cite{Julienne:1998}. This was first achieved by Wynar {\em et
al.}\ \cite{Wynar:2000}, who worked in a Bose-Einstein condensate
of $^{87}$Rb and produced ultracold $^{87}$Rb$_2$ molecules in a
specific vibration-rotation (and hyperfine) state by STIRAP via
the $0_g^-$ state. In their experiment the dump laser was detuned
by only 636 MHz from the pump laser, and the molecules were formed
in the second-to-last vibrationally excited state. Producing
molecules in this way has the major advantage that coherence can
be maintained \cite{Heinzen:2000, Drummond:2002}; this is not the
case if spontaneous emission is involved. Winkler {\em et al.}\
\cite{Winkler:2005} have used 2-colour photoassociation to produce
a coherent superposition of atomic $^{87}$Rb and molecular
$^{87}$Rb$_2$ Bose-Einstein condensates. In a conceptually related
but experimentally quite different approach, Thompson {\em et
al.}\ \cite{Thompson:magres:2005} have used an oscillating {\it
magnetic} field to stimulate deexcitation of atom pairs from free
atom states to molecular states of $^{85}$Rb$_2$.

1-photon and coherent 2-photon photoassociation can be viewed as
manifestations of {\em optical Feshbach resonances}, in which
laser frequencies or intensities are used to tune collision
properties. In a ``dressed state'' picture, the photons of a laser
field bring atomic or molecular states at different energies into
resonance with one another. Such resonances were first
investigated theoretically by Fedichev {\em et al.}\
\cite{Fedichev:1996}, who showed that moderate laser intensities
near resonance could induce sufficient excited-state character to
change scattering lengths and even reverse their sign. Bohn and
Julienne \cite{Bohn:1997} extended this work and introduced the
concept of 2-photon (or 2-colour) optical Feshbach resonances, in
which the state responsible for the resonance is a vibrational
level of the {\it ground} electronic state, connected to the
atomic state by a 2-photon (stimulated Raman) transition.

Optical Feshbach resonances were first demonstrated experimentally
by Fatemi {\em et al.}\ \cite{Fatemi:2000}, who observed changes
in Na-Na scattering properties due to a one-photon resonance as a
function of laser detuning and intensity. Theis {\em et al.}\
\cite{Theis:2004} observed the variation of scattering length
directly, by using Bragg spectroscopy to determine the mean field
energy of an $^{87}$Rb condensate, and Thalhammer {\em et al.}\
\cite{Thalhammer:2005} have carried out similar measurements for a
stimulated Raman resonance.

An optical Feshbach resonance can in principle be used to create
molecules in a very similar way to a magnetic Feshbach resonance.
Javanainen and Mackie \cite{Javanainen:1999} proposed a
photoassociation scheme in which molecules are produced coherently
by chirping (ramping) the laser frequency adiabatically across an
optical Feshbach resonance. Koch {\em et al.}\ \cite{Koch:2005}
have investigated the formation of electronic ground-state
$^{87}$Rb$_2$ molecules in a similar way, and considered ramping
the laser intensity as well as the frequency.

Tuning through optical Feshbach resonances is potentially more
general than magnetic tuning. Laser fields can be switched on and
off much faster than magnetic fields, and optical tuning could be
applied to atoms without nuclear spin, such as the predominant
isotopes of several of the alkaline earths ($^{24}$Mg, $^{40}$Ca,
$^{88}$Sr, $^{138}$Ba). In addition, levels can be tuned into
resonance from much further away using 2-photon resonances than is
possible magnetically. However, a limitation arises because the
extent to which the crossing is avoided depends on the
laser-induced coupling between the two states, and thus on
transition moments and Franck-Condon factors. If the Raman
transition is very weak, it will be impossible to tune across the
resonance slowly enough to achieve adiabatic passage.

\subsection{Coherent control}

Most photoassociation experiments have so far used fixed-frequency
lasers. However, an alternative is to use laser pulses with
tailored frequency and intensity profiles, as has become common in
quantum control experiments on molecules at higher temperature
\cite{Dantus:2004}. A broadband laser can create a non-stationary
state (wavepacket) that is made up of a linear combination of
several different rovibrational levels of the excited electronic
state. The wavepacket then evolves in time, and if carefully
chosen may develop favourable Franck-Condon overlap with low-lying
vibrational levels of the electronic ground state.

\begin{figure} [htbp]
\begin{center}
\includegraphics[width=85mm]{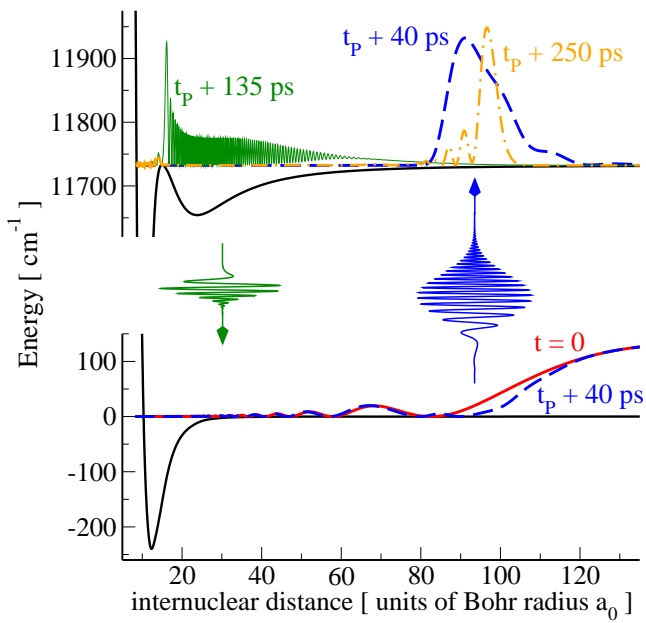}
\caption{Formation of Cs$_2$ using chirped pump and dump pulses,
showing the evolution of the wavepacket on the upper electronic
state ($0_g^-$). Reprinted with permission from Koch {\em et al.}\
\cite{Koch:2006}. Copyright 2006 by the American Physical
Society.} \label{chirp-2colour}
\end{center}
\end{figure}

Vala {\em et al.}\ \cite{Vala:2001} have simulated the use of
chirped picosecond laser pulses to form Cs$_2$ molecules in the
double-minimum $0_g^-$ state, and Luc-Koenig {\em et al.}\
\cite{Luc-Koenig:pra:2004, Luc-Koenig:epd:2004} have investigated
optimization of the pulse characteristics.  Koch {\em et al.}\
\cite{Koch:2006} have simulated two-photon photoassociation using
the scheme shown in Fig.\ \ref{chirp-2colour} and optimized the
parameters of the dump pulse to maximize the formation of
molecules in deeply bound vibrational states. Salzmann {\em et
al.}\ \cite{Salzmann:2006} have carried out initial experimental
work in which evolutionary strategies are used to optimize pulse
parameters to maximise formation of ultracold $^{85}$Rb$_2$, while
Brown {\em et al.}\ \cite{Brown:2006} have found that chirped
femtosecond pulses produce {\it fewer} ultracold $^{85}$Rb$_2$ and
$^{87}$Rb$_2$ molecules than comparable unchirped pulses.

\subsection{Molecules in low vibrational states}

Both photoassociation and magnetic resonance tuning produce
molecules in very high vibrational states. However, molecules in
excited vibrational states can always undergo inelastic collisions
that lead to trap loss. There is thus great interest in finding
ways either to drive the formation of molecules in the vibrational
ground state, $v=0$, or to transfer molecules initially formed in
high-lying states to $v=0$.

Direct photoassociation to form low-lying vibrational states is
not usually feasible for homonuclear molecules, because the
low-energy scattering wavefunction for a pair of atoms has very
little amplitude at short range. There is therefore very little
Franck-Condon overlap with the wavefunctions for low-lying
vibrational states, which are entirely at short range. Various
schemes have been proposed to overcome this \cite{Band:1995,
DeMille:2002, Kotochigova:2004}, but for homonuclear molecules the
combination of parity restrictions and Franck-Condon factors
present formidable obstacles. For example, Jaksch {\em et al.}\
\cite{Jaksch:2002} proposed a 6-photon scheme to form
$^{87}$Rb$_2$ in its ground vibronic state. For heteronuclear
species, on the other hand, the parity restrictions are lifted and
the Franck-Condon factors are more favourable \cite{Wang:1998}, so
that stimulated Raman photoassociation to form $v=0$ molecules may
be feasible \cite{DeMille:2002, Kotochigova:2004}.

\begin{figure} [htbp]
\begin{center}
\includegraphics[width=85mm]{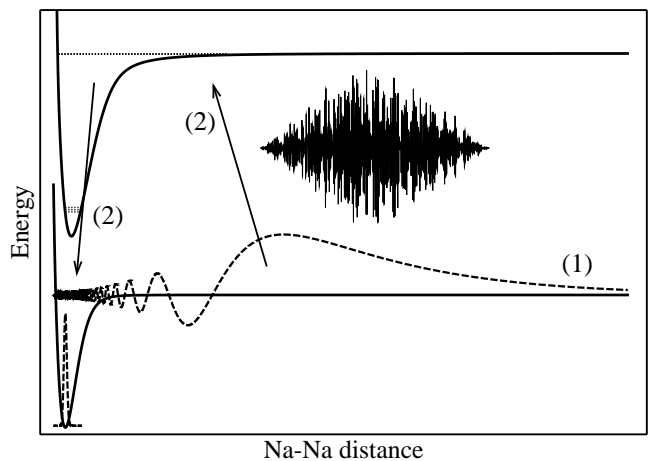}
\caption{The use of tailored laser pulses to stabilise long-range
Na$_2$ molecules by transferring them to the ground vibronic
state. Reprinted with permission from Koch {\em et al.}\
\cite{Koch:2004}. Copyright 2004 by the American Physical
Society.} \label{chirp-deep}
\end{center}
\end{figure}

An alternative approach that may be advantageous is to form
long-range molecules first, by either Feshbach resonance tuning or
photoassociation, and then transfer the molecular population to a
short-range state. Koch {\em et al.}\ \cite{Koch:2004} have used
optimal control theory to design tailored laser pulses that would
achieve this for Na$_2$ as shown in Fig.\ \ref{chirp-deep}, while
Stwalley \cite{Stwalley:2004} has suggested that for heteronuclear
alkali metal dimers it could be efficiently achieved by stimulated
Raman adiabatic passage via mixed levels of the b$^3\Pi$ and
A$^1\Sigma^+$ states.

\begin{figure} [htbp]
\begin{center}
\includegraphics[width=85mm]{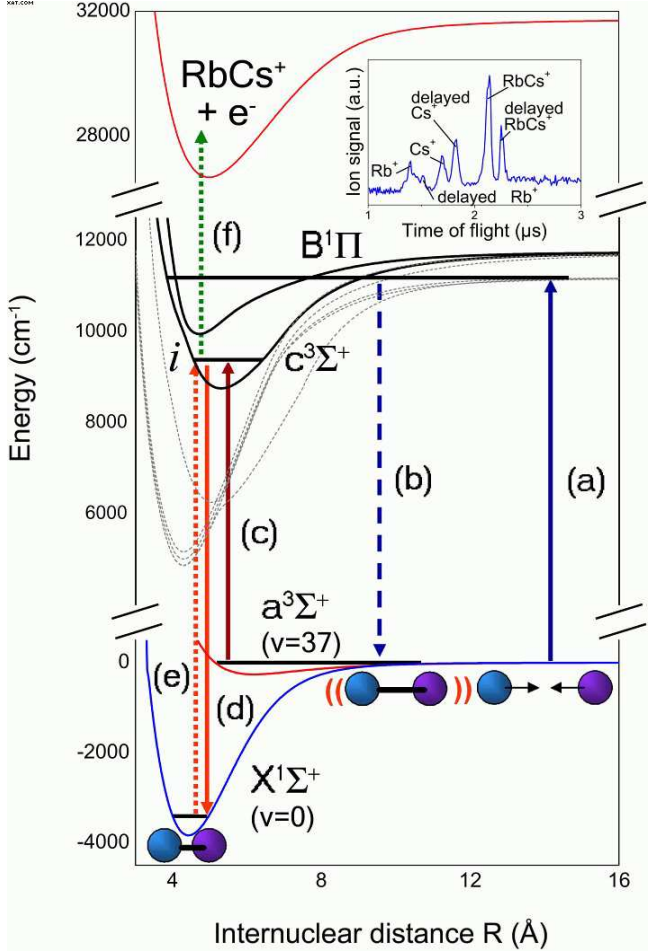}
\caption{The scheme used to produce and detect RbCs molecules in
their ground vibronic state. Reprinted with permission from Sage
{\em et al.}\ \cite{Sage:2005}. Copyright 2005 by the American
Physical Society.} \label{rbcs}
\end{center}
\end{figure}

In the culmination of a series of papers \cite{Kerman:paRbCs:2004,
Kerman:vibdistRbCs:2004, Bergeman:2004}, Sage {\em et al.}\
\cite{Sage:2005} have succeeded in creating ultracold RbCs
molecules ($T\approx100\,\mu$K) in their vibronic ground state
using the 2-step (4-photon) process shown in Fig.\ \ref{rbcs}:
first, molecules are produced in the weakly bound $v=37$ level of
the a$^3\Sigma^+$ state by 1-photon photoassociation (a) followed
by spontaneous emission (b), and then they are transferred to the
$v=0$ or 1 level of the X$^1\Sigma^+$ state by an incoherent
2-photon pump/dump process (stimulated emission pumping, SEP, (c)
and (d) in Fig.\ \ref{rbcs}) via a mixed level of the $c$ and $B$
excited states. The SEP process has an efficiency of only 6\%, but
this could in principle be dramatically improved by using STIRAP
instead.

Other heteronuclear molecules such as KRb \cite{Mancini:2004,
Wang:KRb-EPD:2004, Wang:2005} and NaCs \cite{Haimberger:2004} have
also been produced in the electronic ground state
\cite{Mancini:2004, Wang:KRb-EPD:2004, Haimberger:2004} and
state-selectively detected \cite{Wang:2005}, but not yet
transferred to low-lying vibrational states. LiCs and NaCs have
been formed in the lowest vibrational level of the lowest triplet
state on the surface of helium droplets \cite{Mudrich:2004}, but
the temperature is that of the droplet (0.38\,K) and cannot easily
be lowered further.

\section{Molecules in optical lattices}

An optical lattice is formed by standing waves between two or more
laser beams. Since atoms and molecules are polarisable, they
experience a periodic potential with minima at the points where
the electric fields due to the laser are greatest. The separation
between successive minima is half the laser wavelength, and the
heights of the barriers between minima can be adjusted by varying
the laser intensity. Confinement of atoms in optical lattice cells
was first observed by Westbrook {\em et al.}\
\cite{Westbrook:1990}. One particularly interesting state that can
be created is a Mott insulator phase
\cite{Fisher:1989,Greiner:2002}, in which the lattice sites are
occupied in a regular pattern and which can be ``melted" to form a
superfluid by lowering the barriers. The dynamics of Bose-Einstein
condensates in optical lattices have been recently reviewed by
Morsch and Oberthaler \cite{Morsch:2006}.

\begin{figure} [htbp]
\begin{center}
\includegraphics[width=85mm]{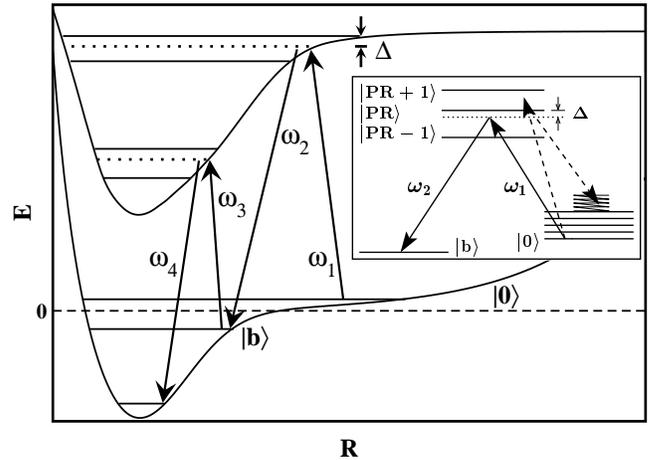}
\caption{Scheme for creating molecules at doubly occupied lattice
sites in a Mott insulator. Note the quadratic trapping potential
at long range that serves to confine the initial state. Reprinted
with permission from Jaksch {\em et al.}\ \cite{Jaksch:2002}.
Copyright 2002 by the American Physical Society.}
\label{mol-lattice}
\end{center}
\end{figure}

The production of molecules in optical lattices offers intriguing
possibilities. The barriers between lattice sites enhance
stability by preventing collisions between molecules at different
sites. Jaksch {\em et al.}\ \cite{Jaksch:2002} proposed creating a
molecular Bose-Einstein condensate by Raman photoassociation in a
Mott insulator with 2 atoms in each lattice site as shown in Fig.\
\ref{mol-lattice}. The trapping potential discretizes the
molecular continuum and converts the free-bound photoassociation
process into a bound-bound transition. Damski {\em et al.}\
\cite{Damski:2003} extended this idea to the creation of a dipolar
superfluid by photoassociation in a lattice containing one atom of
each of two different species in each site. Rom {\em et al.}\
\cite{Rom:2004} have created $^{87}$Rb$_2$ molecules in an optical
lattice by essentially the technique of ref.\
\onlinecite{Jaksch:2002}. Ryu {\em et al.}\ \cite{Ryu:2006} have
carried out similar experiments and observed coherent oscillations
(Rabi cycling) between atomic and molecular gases. They also
observed distinct Raman spectra for photoassociation occurring in
doubly and triply occupied sites. Thalhammer {\em et al.}\
\cite{Thalhammer:2006} have also created $^{87}$Rb$_2$ on an
optical lattice, but by magnetic tuning through a Feshbach
resonance. They were able to purify the resulting molecular
lattice by driving out the remaining atoms with a laser, and
showed that the molecules remained trapped much longer (700 ms)
after the atoms had been removed than while they were still
present. They attribute the loss rate to inelastic atom-molecule
collisions that occur when atoms tunnel through lattice barriers
into sites occupied by molecules; once the atoms are removed, the
corresponding loss due to molecule-molecule collisions is much
slower simply because of the larger mass and reduced tunnelling
rate of the molecules. Very recently, Winkler {\em et al.}\
\cite{Winkler:2006} have created long-lived bound atom pairs with
a {\it repulsive} atom-atom interaction. The pairs cannot decay
into separated atoms because the energies of the atoms that would
be produced are not allowed by the lattice band structure.
Remarkably, the pairs {\it do} decay if the repulsive interaction
between the atoms is switched off by tuning the scattering length
to zero. Volz {\em et al.}\ \cite{Volz:2006} have created a Mott
state of $^{87}$Rb$_2$ molecules and demonstrated that phase
coherence is restored when the lattice depth is reduced.

Initial experiments have also been carried out on fermion dimers
in optical lattices. Moritz {\em et al.}\ \cite{Moritz:2005}
reported the creation of a 1-dimensional gas of $^{40}$K$_2$
molecules in a 2-d optical lattice by Feshbach resonance tuning,
and St\"oferle {\em et al.}\ \cite{Stoferle:2006} have carried out
similar experiments in a 3-d lattice. However, the tunnelling
rates for the lattices used were too fast for lifetimes to be
increased beyond those normal for fermion dimers.

Optical lattices provide a very promising environment for bringing
together more than two atoms at a time and studying many-body
processes under controlled conditions. Stoll and K\"ohler
\cite{Stoll:2005} have suggested a scheme that could be used to
produce Efimov states of alkali metal trimers directly from 3
atoms by magnetic Feshbach resonance tuning in an optical lattice.

\section{Collisions of ultracold molecules}

Collision processes involving ultracold molecules are of prime
importance to trapping  and controlling them. As described above,
inelastic collisions usually release enough kinetic energy that
both collision partners are lost from the trap. The initial
experiments on boson dimers formed by Feshbach resonance tuning
\cite{Donley:2002, Herbig:2003, Xu:2003, Durr:mol87Rb:2004} gave
lifetimes that suggested vibrational relaxation rates for
atom-molecule collisions around $10^{-10}$ cm$^3$ s$^{-1}$.
Quantitative estimations were not usually attempted, but this was
consistent with the rate of $1.6\times10^{-10}$ cm$^3$ s$^{-1}$,
estimated by Yurovsky {\em et al.}\ \cite{Yurovsky:2000} on the
basis of early Feshbach resonance experiments \cite{Inouye:1998,
Stenger:1999}. It also agreed with quantum dynamics calculations
by Soldan {\em et al.}\ \cite{Soldan:2002} on vibrational
relaxation in Na + Na$_2$ collisions. More recently, Mukaiyama
{\em et al.}\ \cite{Mukaiyama:2004} have measured the trap loss
rate for $^{23}$Na$_2$ molecules formed by Feshbach resonance
tuning and obtained an atom-molecule rate coefficient $k_{\rm
loss} = 5.1\times10^{-11}$ cm$^3$ s$^{-1}$ for molecules in the
highest vibrational state.

Relaxation processes involving molecules formed by
photoassociation have also been studied. Wynar {\em et al.}\
\cite{Wynar:2000} obtained an upper bound of $k_{\rm
loss}=8\times10^{-11}$ cm$^3$ s$^{-1}$ for $^{87}$Rb$_2$ molecules
in the second-to-last vibrationally excited state. Staanum {\em et
al.}\ \cite{Staanum:2006} have investigated inelastic collisions
of rovibrationally excited Cs$_2$ ($^3\Sigma_u^+$) in collisions
with Cs atoms in two different ranges of the vibrational quantum
number $v$ by monitoring trap loss of Cs$_2$. They obtained
atom-molecule rate coefficients close to $1.0\times10^{-10}$
cm$^3$ s$^{-1}$ for both $v=4$ to 6 and $v=32$ to 47. Zahzam {\em
et al.}\ \cite{Zahzam:2006} have carried out similar work for
different rovibrational states of $^3\Sigma_u^+$, but also
considered molecules in the $^1\Sigma_g^+$ state and
molecule-molecule collisions. They obtained rate coefficients of
$2.6\times10^{-11}$ cm$^3$ s$^{-1}$ and $1.0\times10^{-11}$ cm$^3$
s$^{-1}$ in the atom-atom and atom-molecule cases respectively,
both with quite large error bounds.

\begin{figure}
\begin{center}
\includegraphics[angle=-90,width=85mm]{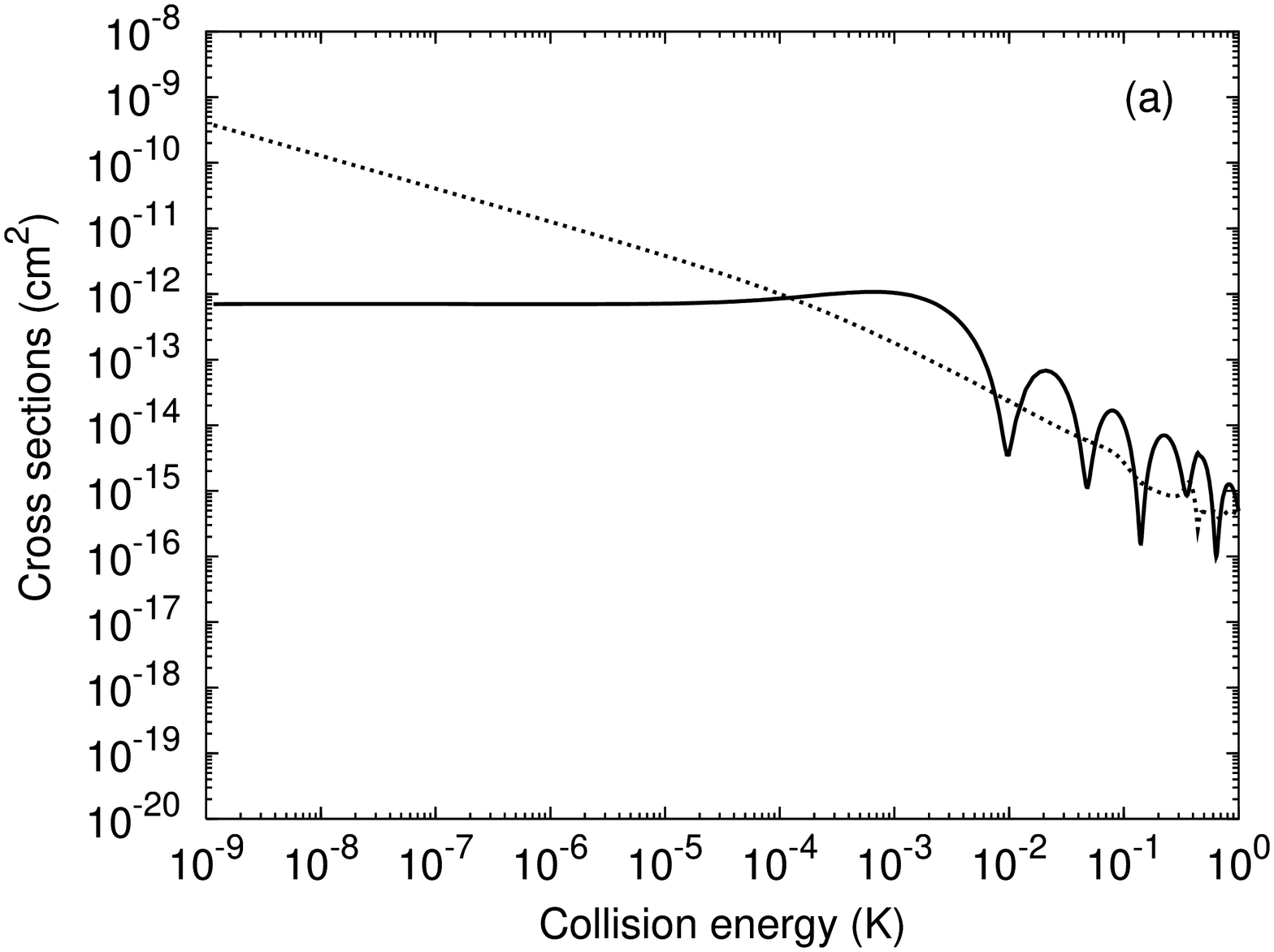}
\includegraphics[angle=-90,width=85mm]{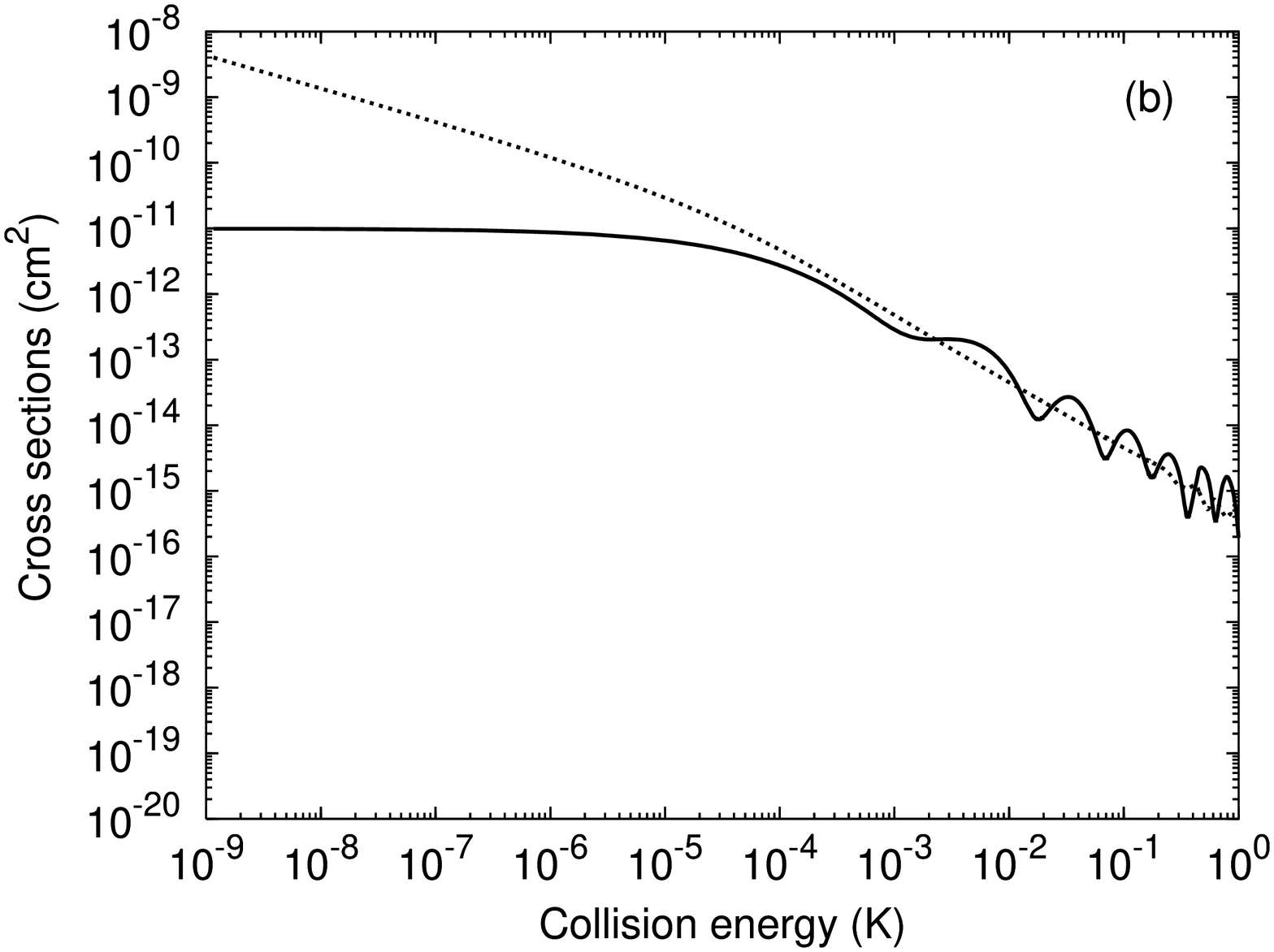}
\end{center}
\caption{ Cross sections from s-wave quantum reactive scattering
calculations for Na + Na$_2$($v=1,j=0$) Elastic and quenching
results are shown as solid and dotted lines. The cross sections on
the additive potential (a) are a factor of 10 smaller than those
on the non-additive potential (b). Reproduced from Sold\'{a}n {\em
et al.}\ \cite{Soldan:2002}} \label{figna}
\end{figure}

Sold\'{a}n {\em et al.}\ \cite{Soldan:2002}, Qu\'em\'ener {\em et
al.}\ \cite{Quemener:2004, Quemener:2005}, and Cvita\v s {\em et
al.}\ \cite{Cvitas:bosefermi:2005, Cvitas:li3:2006,
Cvitas:hetero:2005} have carried out quantum dynamics calculations
on atom-molecule collisions between alkali-metal atoms and dimers.
This work will be described in greater detail in a forthcoming
review \cite{Soldan:review:2007}. The calculations used a reactive
scattering approach developed by Launay and LeDourneuf
\cite{Launay:1989}, which has been applied extensively to chemical
reactions such as N($^2$D)+H$_2$ \cite{Honvault:1999} and
O($^1$D)+H$_2$ \cite{Honvault:2001} at higher energies. Sold\'{a}n
{\em et al.}\ \cite{Soldan:2002} showed that {\it barrierless atom
exchange reactions} can occur in Na + Na$_{2}$, and that even at
very low energy such collisions cause very fast vibrational
relaxation ($k_{\rm inel}$ on the order of $10^{-10}$ cm$^3$
s$^{-1}$) for collisions of molecules in low vibrational states.
The cross sections were shown to depend strongly on the details of
the potential energy surfaces, and to change by a factor of 10
when non-additive forces were included, as shown in Fig.\
\ref{figna}. In subsequent work, Sold\'{a}n {\em et al.}\
\cite{Soldan:2003} showed that non-additive forces are important
for all the alkali metal trimer systems, and affect the well depth
for spin-polarized Li + Li$_2$ collisions by a factor of 4. The
quantum dynamics calculations were subsequently extended to Li +
Li$_2$ collisions, both isotopically homonuclear
\cite{Cvitas:bosefermi:2005, Cvitas:li3:2006} and heteronuclear
\cite{Cvitas:hetero:2005}, and to K + K$_{2}$
\cite{Quemener:2005}. For the homonuclear Li systems, Cvita\v s
{\em et al.}\ \cite{Cvitas:bosefermi:2005} demonstrated that there
is {\it no} systematic suppression of the inelastic collision
rates for fermion dimers in low vibrational states, in contrast to
the situation for molecules in long-range states
\cite{Strecker:2003, Cubizolles:2003, Jochim:Li2pure:2003,
Regal:lifetime:2004, Petrov:2004, Petrov:suppress:2005}. This is
very important for attempts to transfer molecules formed by
Feshbach resonance tuning to the vibrational ground state, because
it means that the transfer must be accomplished {\it without}
spending significant time in intermediate vibrational states.

\section{Conclusions}

There have been enormous advances in the production and
manipulation of molecules in laser-cooled atomic gases. Molecules
have been produced both by photoassociation and by magnetic tuning
through Feshbach resonances. Molecular Bose-Einstein condensates
have been produced for molecules in long-range states, and the
first signatures of triatomic and tetraatomic molecules have been
seen. Most of the experimental advances were guided by theoretical
predictions, and the experiments have in turn stimulated a large
number of theoretical studies.

Prospects for the future include the use of cold molecules for
high-precision measurement and the production of
quantum-degenerate gases of ground-state molecules, which will be
stable to collisions and offer a wealth of new possibilities for
quantum control. Heteronuclear molecules are particularly
interesting, because they can have substantial dipole moments in
long-range states. Dipolar quantum gases offer a new range of
novel properties, and ultracold polar molecules also have
potential applications in quantum computing and in studying
fundamental physical properties such as parity violation and the
electron dipole moment.

\section{Acknowledgments}

The authors are grateful to Simon Cornish and Charles Adams for
valuable discussions and to Eite Tiesinga and Cheng Chin for
providing Figures based on unpublished work. PS acknowledges
support from the Ministry of Education of the Czech Republic
(grant no. LC06002).

%\bibliography{cold1,cold2,cold3,cold4,cold5,cold6,cold7,dimerspec,molpro,react,trap}
%\end{document}

\end{document}